\documentclass{article}

\usepackage{arxiv}

\usepackage[utf8]{inputenc} % allow utf-8 input
\usepackage[T1]{fontenc}    % use 8-bit T1 fonts
\usepackage{hyperref}       % hyperlinks
\usepackage{url}            % simple URL typesetting
\usepackage{booktabs}       % professional-quality tables
\usepackage{amsfonts}       % blackboard math symbols
\usepackage{nicefrac}       % compact symbols for 1/2, etc.
\usepackage{microtype}      % microtypography
\usepackage{lipsum}		% Can be removed after putting your text content
\usepackage{graphicx}
\usepackage{natbib}
\usepackage{doi}

\usepackage{mathtools}

\graphicspath{ {./figures/} }

\title{When regression coefficients change over time: A proposal}

\author{ \href{https://orcid.org/0000-0003-4058-1543}{\includegraphics[scale=0.06]{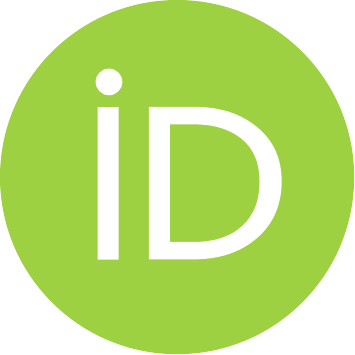}\hspace{1mm}Malte~Schierholz}\thanks{This work was supported by a postdoc fellowship of the German Academic Exchange Service (DAAD). It was created during a research stay at the Machine Learning for Public Policy lab at Carnegie Mellon University.} \\
	Department of Statistics\\
	Ludwig-Maximilians-Universität München\\
	80539 Munich \\
	\texttt{malte.schierholz@stat.uni-muenchen.de} \\
}

\renewcommand{\shorttitle}{Coefficients that change over time}

\hypersetup{
pdftitle={When regression coefficients change over time: A proposal},
pdfauthor={Malte~Schierholz},
pdfkeywords={Distribution shift, Forecasting individual outcomes, Regression, State space methodology}
}

\begin{document}
\maketitle

\begin{abstract}
	A common approach in forecasting problems is to estimate a least-squares regression (or other statistical learning models) from past data, which is then applied to predict future outcomes. An underlying assumption is that the same correlations that were observed in the past still hold for the future. We propose a model for situations when this assumption is not met: adopting methods from the state space literature, we model how regression coefficients change over time. Our approach can shed light on the large uncertainties associated with forecasting the future, and how much of this is due to changing dynamics of the past. Our simulation study shows that accurate estimates are obtained when the outcome is continuous, but the procedure fails for binary outcomes.
\end{abstract}

% keywords can be removed

\section{Introduction}\label{sec:introduction}

Time is a crucial factor in many data sets, but researchers do not have agreed upon best practices of how date variables should be used.

For example, consider the DonorsChoose data set.\footnote{The KDD Cup 2014 DonorsChoose dataset is available from \href{https://www.kaggle.com/c/kdd-cup-2014-predicting-excitement-at-donors-choose/data}{Kaggle}. The \href{https://github.com/dssg/donors-choose}{Data Science for Social Good Github Repository} lays out a possible route of analysis, our reference method here.} DonorsChoose (\url{https://www.donorschoose.org/}) allows teachers to crowdsource funding for various projects, e.g., funding to buy a microwave that lets them cook with children in preschool. The requested items are shipped only when enough citizen donors support the project, and funding goals are met within four months. As data scientists, we want to identify postings automatically that are at risk of not meeting its funding goal, and before they get published, with the intention to help teachers improve their postings and, thereby, their chances to get funded. Several projects are posted each day. It is not unlikely that factors affecting the success of funding change over time.

A common framework to predict outcomes in settings like DonorsChoose is implemented in triage\footnote{\url{https://pypi.org/project/triage/}}, a python package. The temporal cross-validation framework is as follows: 
\begin{itemize}
    \item \textit{Training data}: Pool all new postings from a given month, forming the training data. Include features as they were available at the time of the posting. The outcome, funding success after four months, is added, although it only becomes available with a four-month delay. In general, for applications other than DonorsChoose the interval over which observations are pooled and the delay period will of course differ.
    \item \textit{Model training}: Employ any statistical learning technique (e.g., random forest, boosting, linear regression, ...) to generate a predictive model. The necessary data are only available after the four-month delay.
    \item \textit{Deployment/Test data}: The predictive model gets deployed to predict funding success in future time periods. It is good practice to evaluate it first. For evaluation, all postings are pooled from the month (or another period) when the model would get deployed, forming the test data.
    \item \textit{Check sensitivity}: Repeat the steps above for different training/test months (or other periods), while keeping the time period between both data sets constant. If the performance varies over time, this is a reason to worry.
\end{itemize}

Temporal cross-validation is in a sense self-contradictory. While it acknowledges that correlations found in the training data may differ in the test data, the same is not true for possible changes that may occur during the training period (or during the test period). Instead, the pooled training (or test) cases are usually analyzed as if they were collected on a single day; the time dimension within the training (or test) data gets lost.

Besides, what is the optimal length of the interval for pooling the training/test data when the dynamics shift over time? How much shift is to be expected happening after the training features were generated but before the model can get deployed? Statistical learning theory provides little guidance here, since a standard theoretical assumption---training and test data are drawn from the same probability distribution---is not met. Instead, the joint distribution of features and outcome, $p_t(y, x)$, might shift over time. In the most extreme case, if there were a sudden shock between training and testing and the dynamics change in unforeseeable ways, extrapolation from the training to the deployment period would be impossible. The contrary case, that the distribution $p_t(y, x)$ is constant in time, would allow ideal predictions, but real data rarely behave this way (as one easily sees with the sensitivity checks mentions above). The middle ground is most realistic, namely that $p_t(y, x)$ is a continuous function over time that does change, but not too much. While researchers have proposed various ways to correct for distribution shift \citep[e.g.,][pp. 133]{kim_universal_2022, varshney_trustworthy_2021}, we are not aware of any research that aims to model the size of such shifts, while assuming continuity.

In our fully Bayesian approach, we do not pool training data, but model changing parameters at consecutive time points, adopting techniques from the state space literature \citep[e.g.,][]{durbin_time_2012}. This is advantageous for three reasons. First, we can generate credibility intervals describing the uncertainty when predicting future outcomes. Second, our approach automatically detects possible long-term trends that would easily be missed within the standard framework mentioned above. Third, shocks that occur during the training period in the standard framework can have negative consequences for model performance (using training data from before the shock bodes ill for extrapolation into the future). Our approach does not average over observations from before and after the shock, but adapts automatically to such changing realities, only perpetuating the most recent dynamics from after the shock.

This paper proceeds as follows. Section 2 describes our approach. Sections 3 and 4 showcase simulation studies, illustrating the models behavior. Section 5 concludes.

\section{Model}\label{sec:model}

Let the data have $i = 1, ..., N$ observations. We want to predict an outcome $y$ given predictors $x$. $y$ will be continuous in the beginning, but we will also test binary outcomes in later simulations. While this situation can be modeled by least-squares regression (or any other regression algorithm), our situation differs because we have information about time: The predictors about a given individual are available at time $t \in \{1, ..., T\}$, the point in time at which we would like to make the forecast (which can differ from the time of measurement). The outcome will be observed later, within a fixed time span (if the outcome were already known, we would not need to predict it).

Let $y_{i(t)}$ be the outcome of individual $i$, observed within a fixed period after $t$ (not yet available at time $t$, although the notation suggests so!). $x_{i(t)p}$ are the $p = 1, ..., P$ predictors that were available at the time of prediction $t$. $\beta$, $\alpha$ and $\nu$ are parameter matrices, each of dimension $T \times P$, with entries for each predictor at all time points.

Our modeling approach is described by the following equations:
\begin{eqnarray}
y_{i(t)} | x_{i(t)p} \sim N(\sum_p x_{i(t)p} \beta_{tp}, \sigma_y^2) \textrm{~~~~~~~~~~~~~~~} & \textrm{(observation equation)} \\
\beta_{tp} \sim N(\alpha_{tp}, \sigma^2_{\beta}) \textrm{~~~~~~~~~~~~~~~} \forall t,p \textrm{~~~~~~~} & \textrm{(fluctuation equation)} \\
\alpha_{t+1,p} \sim N(\alpha_{t,p} + \nu_{t,p}, \sigma^2_{\alpha}) \textrm{~~~} \forall t,p \textrm{~~~~~~~} & \textrm{(state equation)} \\
\nu_{t+1,p} \sim N(\nu_{t,p}, \sigma^2_{\eta})\textrm{~~~~~~~~~~~~~~~} \forall t,p\textrm{~~~~~~~} & \textrm{(trend equation, optional)}
\end{eqnarray}

Without prior knowledge, we use non-informative priors for the variances $\sigma_y^2$, $\sigma^2_{\beta}$, $\sigma^2_{\alpha}, \sigma^2_{\eta}$ and for the vectors $\alpha_{1,p}$ and $\nu_{1,p}$ at $t=1$.

The observation equation would be a least-squares regression if it were not for the subscript $t$, indicating that parameters $\beta$ vary and depend on the hour/day/year (application-specific) of prediction. There may be very few observations at each time point, impeding the estimation of $\beta$. Since state-space models (also known as dynamic linear models) \citep{prado_time_2010, shumway_time_2011, durbin_time_2012} were developed for situations, in which a single observation (or none) is available at each $t$, they appear well-suited for our analysis. The state equation and the trend equation are well-known concepts in that literature, and many extensions, for example to include seasonal patterns in the model, are available.

Equations (2)-(4) denote independent draws from univariate Gaussians for each of the $P$ predictors. Generalizations using multivariate Gaussians, which would allow for prior correlations $\neq 0$ between coefficients, should be possible as well, but have not been implemented yet.

Unlike other approaches we are aware of, we include a fluctuation equation, only possible because we usually have (far) more than one observation at each time point. Standard state-space models would model $y_i$ (observation equation) in direct dependence of $\alpha_t$ (state equation). The implication is this: If we had at each time point an infinite number of observations, $\sigma^2_\beta$ would go to $\infty$ and the $\beta_{tp}$ would be equal to estimates of $t$ independent regressions. If there are, conversely, just very few observations at each time point, $\sigma^2_\beta$ will turn out to be small and observations from neighboring time points weigh in strongly in the estimation of $\beta_{tp}$.

Our approach is purely Bayesian. We estimate parameters using the probabilistic modeling language Stan \citep{stan_development_team_rstan_2021}, see supplementary files online.

\section{Simulation Study for Continuous Outcomes}

\begin{figure}
	\centering
	\includegraphics[width=630pt, angle=270]{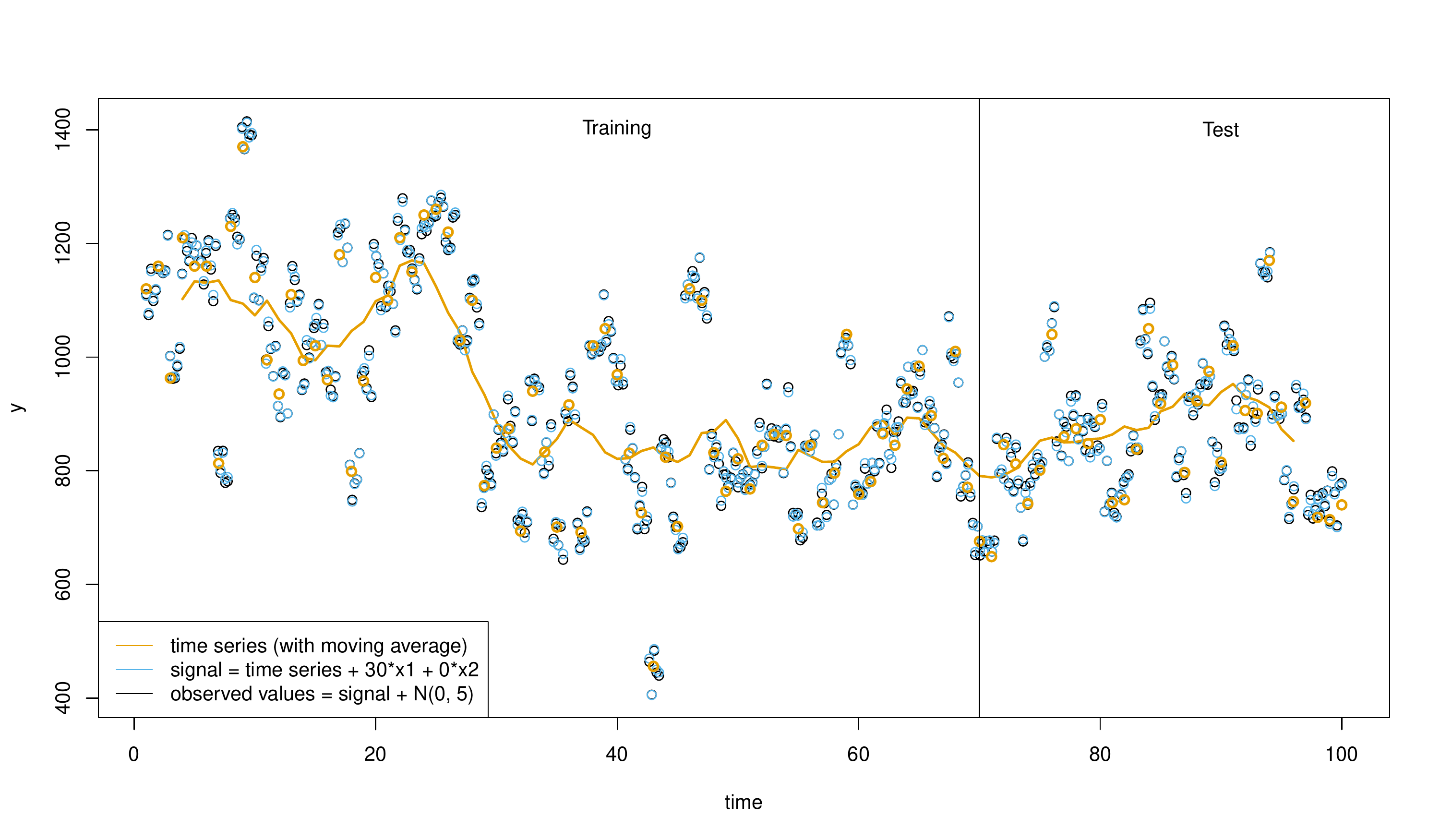}
	\caption{Overview of simulated data.}
	\label{fig:fig1}
\end{figure}

To test the estimation procedure, we run a simulation study. We start with a time series, $ts_t$, containing 100 time points, shown in yellow in Figure \ref{fig:fig1} (the Nile data from \citep{durbin_time_2012}). $ts_t$ is directly related to $\alpha_{t,0}$, which is to be interpreted as the smoothed version of this time series. Our simulated data have five observations at each time point. The covariates $x_{i1} \sim N(0,1)$ and $x_{i2} \sim N(0, 20)$ are drawn from two independent normal distributions. The observed $y$-values (the black dots in Figure \ref{fig:fig1}) are then generated using the formula

\begin{align}
y_{i(t)} = ts_t + 30 \cdot x_{i1} + 0 \cdot x_{i2} + N(0, 5)    
\end{align}

Blue dots in Figure 1 depict the outcome values before adding the noise $N(0, 5)$. They are only shown for illustration (noise is very small in our setup!) and are not used for estimation.

The goal of the simulation is to see whether we can recover the original time series and the parameters $\beta_{1t} = 30$ and $\beta_{2t} = 0$ using observed data $x_{i(t)1}$, $x_{i(t)2}$ and $y_{i(t)}$ only. Only observations from $t = 1, ..., 70$ are used for model training, the final 30 time points are used for evaluation. The delay (time between availability of features and outcomes) in this simulation is zero, otherwise there would be a gap between training and test data. Our evaluation showcases for certain parameters and predictions of interest the posterior means from 8000 MCMC draws, along with their .95-credibility intervals.

\begin{figure}
	\centering
	\includegraphics[width=\textwidth]{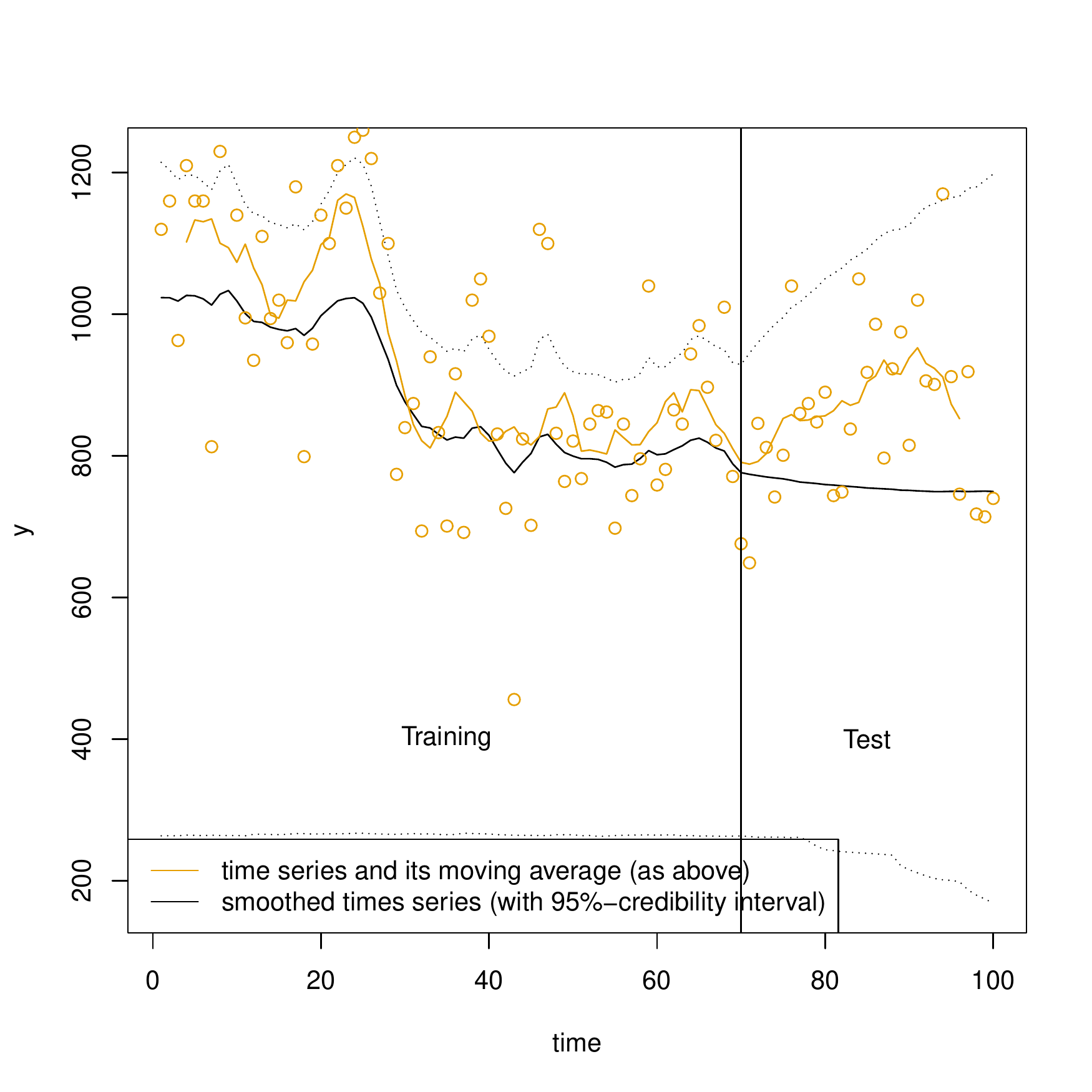}
	\caption{Compare smoothed and true time series over time.}
	\label{fig:fig2}
\end{figure}

Figure \ref{fig:fig2} shows the smoothed time series from our model (in black) against the original time series $ts_t$ (in yellow). During the training phase, the smoothed time series closely aligns to the moving average, showing that the model is capable to detect changes over time. This time series exhibits strong changes over time, and for this reason it is hard make predictions of how the series will behave during the test period, as evidenced by the widening credibility intervals. At the end of the training phase the model picks up a downward slope from the data, and this trend is remembered during the test phase, leading to the slight decline. If we had not included the optional trend component in our model, the latest known state from the training phase would have been predicted (=a horizontal black line during the test phase).

\begin{figure}
	\centering
	\includegraphics[width=\textwidth]{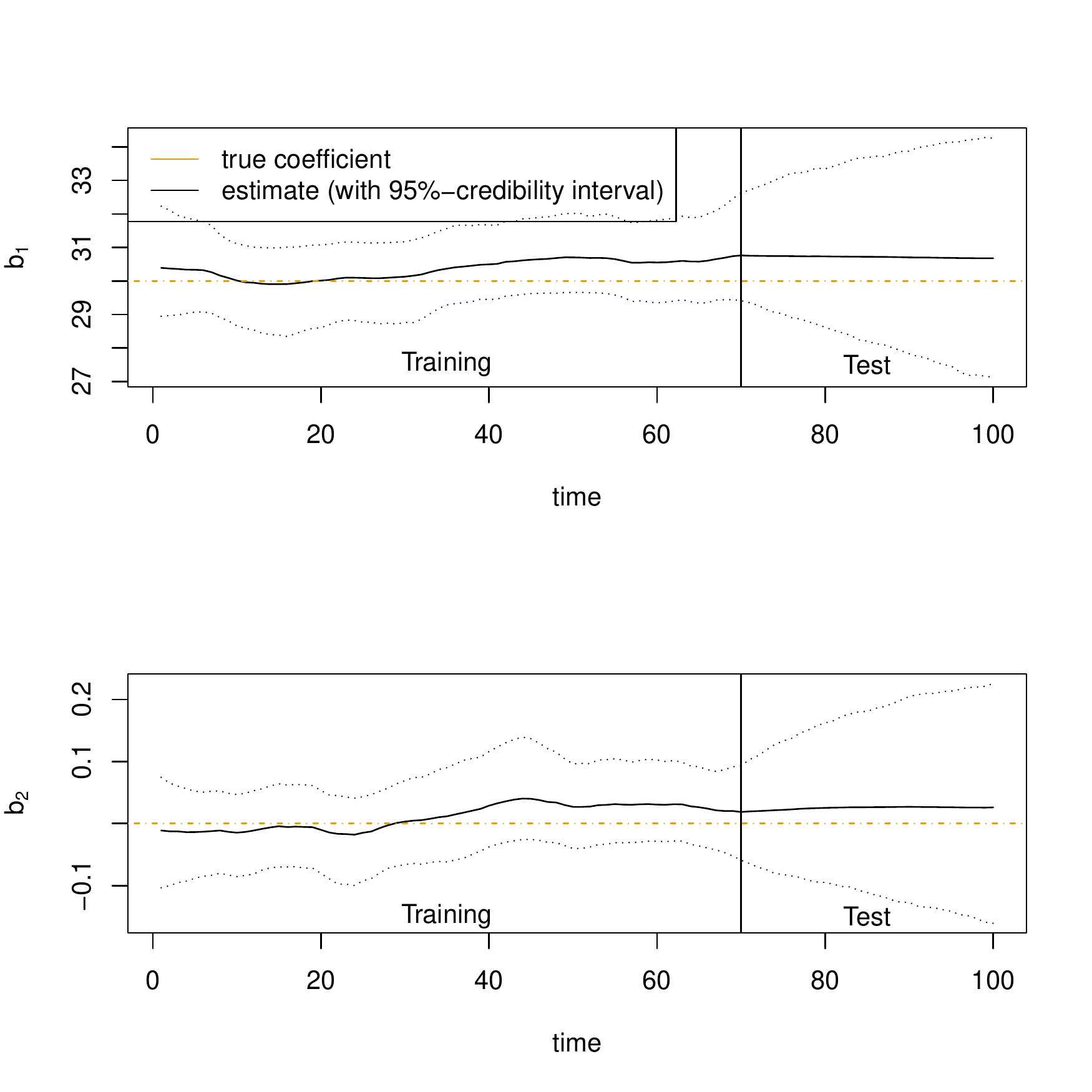}
	\caption{Compare estimated and true coefficients $\beta_{1t}$ and $\beta_{2t}$ over time}
	\label{fig:fig3}
\end{figure}

Figure \ref{fig:fig3} shows the estimated parameters for $\beta_{1t}$ and $\beta_{2t}$. The true values (known from the simulation setup) are always within the credibility intervals, despite very narrow intervals.

\begin{figure}
	\centering
	\includegraphics[width=\textwidth]{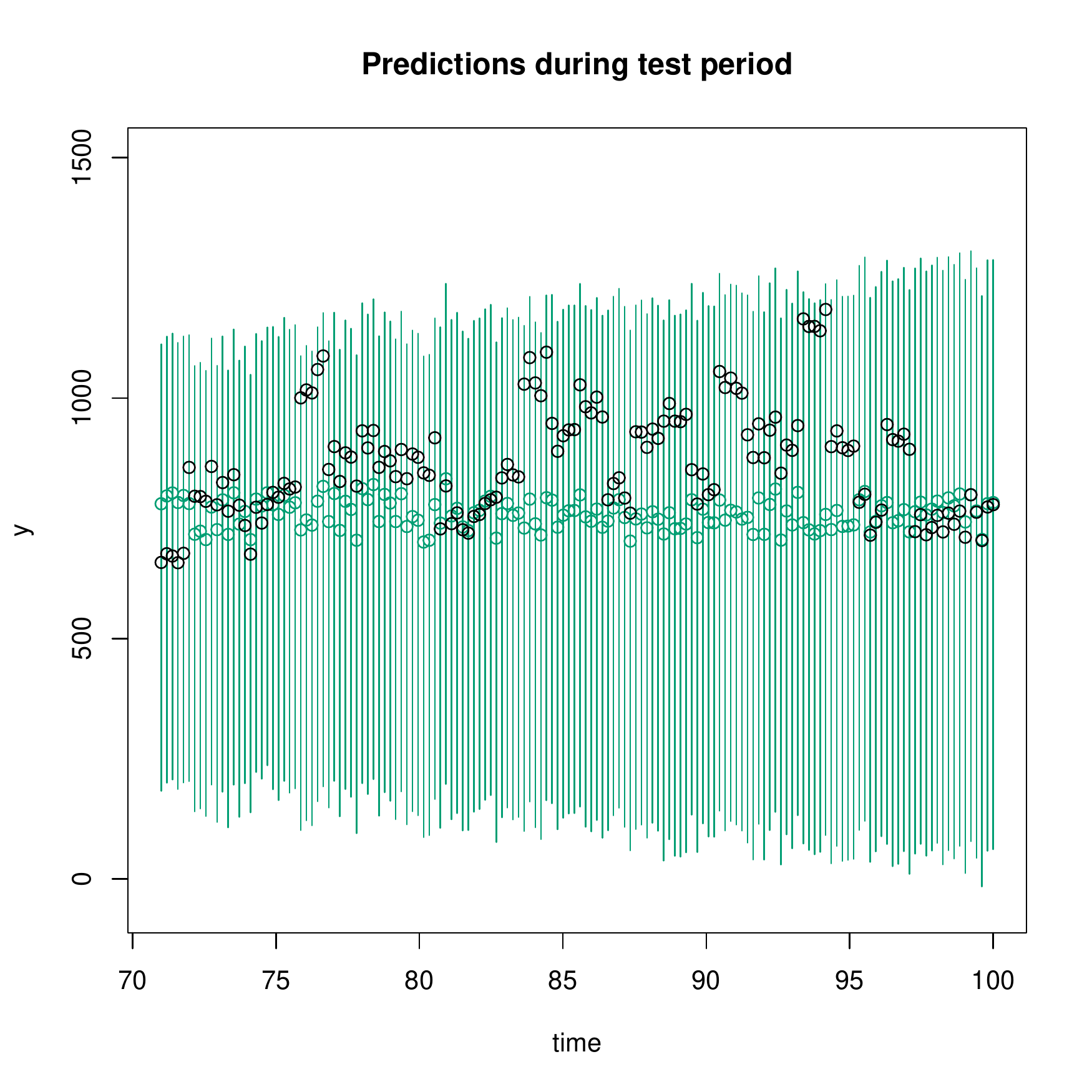}
	\caption{Compare predicted values and true values during the test period. Predicted values (with .95-credibility intervals) are shown in green; True values in black.}
	\label{fig:fig4}
\end{figure}

Using these estimates, we can make predictions for individual test cases (see Figure \ref{fig:fig4}). Because the model could not foresee the upward trend in the time series that only occurred after the training period ended, the true values are generally equal to or larger than the predicted values. Though, the true values are all within their credibility intervals, showing that this model is good in quantifying its uncertainty.

\section{Simulations for Binary Outcomes}\label{sec:binary}

To model binary outcomes, the observation equation in (1) can easily be changed to a logistic regression,

\begin{eqnarray}
y^{(bin)}_{i(t)} | x_{i(t)p} \sim Bin(1, \textrm{logit}^{-1}(\sum_p x_{i(t)p} \beta_{tp})) & \textrm{(observation equation)}
\end{eqnarray}
without making other changes.

However, meaningful results as shown for continuous outcomes above cannot be replicated with binary outcomes. In three simulations we keep the setup for continuous outcomes, except that we draw binary outcomes now. In particular, we 

\begin{enumerate}
    \item draw the outcome $y^{(bin)}_{i(t)}$ from a Bernoulli distribution with probability $p = \textrm{logit}^{-1}(ts_t - \bar{ts_t} + 30 \cdot x_{i1} + 0 \cdot x_{i2})$
    \item set $y^{(bin)}_{i(t)}$ to 1 if $y_{i(t)} > ts_t + 30 \cdot x_{i1} + 0 \cdot x_{i2}$
    \item set $y^{(bin)}_{i(t)}$ to 1 if $y_{i(t)} > 900$
\end{enumerate}

For the third simulation, the MCMC-chains do not mix, implying non-convergence, and we are unable to determine the posterior distribution. For the first two simulations, we also obtain warnings that chains do not mix, but preliminary graphical checks indicate that sufficient mixing occurred. Yet, the estimates do not resemble or correlate with the true values from the simulation setup in any meaningful way.

The model was also tested with the DonorsChoose dataset, but the MCMC chains did not mix.

\section{Discussion}\label{sec:discussion}

We propose a new approach for time-related data analysis, adopting concepts from the state space literature. This approach allows explicit modeling of how coefficients in least-square regressions change over time. Our simulation provides promising results for continuous outcomes.

In which situations would this approach be most useful? We do not recommend using it if the regression coefficients of the available data do not change over time (statistical learning is probably the better method here), but for situations when coefficients do change. Abrupt changes of parameters, e.g., due to economic shocks, are not explicitly modeled. The model should still be robust against such abrupt changes when they happen once or twice during the training period, but has no way to account for unexpected future shocks.

This research started with the goal to apply this approach with binary outcomes and logistic regression, and this was not successful. To move forward, we recommend further simulation studies. For binary outcomes, we wonder under which circumstances one could still make progress, or if theoretical reasons make this impossible. For continuous outcomes, we argued that our approach should be most fruitful in the presence of long-term trends or sudden shocks, but this reasoning should be confirmed in simulation studies. Other simulations should explore issues such as the following: What is the minimal number of time points needed for our approach and what else can be learned from such simulations of reduced complexity? When using simulated data from the proposed model (and not using the 100 Nile data points we have used here), can additional, higher-level model parameters be estimated consistently? Can the parameters still be estimated if the true parameters fluctuate over time?

Our observation equation, least-squares regression, assumes linearity of parameters, making it rather restrictive. Interactions would need to be hand-coded, and allowing for arbitrary functional forms is rather challenging. Conversely, statistical learning algorithms allow the estimation of more flexible functional forms and automated feature selection. This promises better performance and makes the feature engineering processes more efficient, so that statistical learning is the preferred option for many. Statistical learning that takes time into account would, for example, be possible using Long Short-Term Memory (LSTM) Artificial Neural Networks \citep{hochreiter_long_1997}. A comparison with our approach would be valuable.

Maybe the most important feature of our approach is being explicit on prognostic uncertainty, while allowing for shifting distributions over time. Our experiences are already promising, but more research will be needed.

\bibliographystyle{unsrtnat}
\bibliography{references2}  %%% Uncomment this line and comment out the ``thebibliography'' section below to use the external .bib file (using bibtex) .





% \end{thebibliography}

\end{document}